\documentclass[prb,twocolumn,showpacs,preprintnumbers,amsmath,amssymb]{revtex4}
\usepackage{amssymb}
\usepackage{bm}
\usepackage{amsmath}
\usepackage[dvips]{graphicx}
\usepackage{color}

\begin{document}

\title{
Binding of a mobile hole by an impurity potential in the
t-J model: parity breaking
}

\author{O. P. Sushkov}
\affiliation{School of Physics, University of New South Wales, Sydney 2052, Australia}
\author{J. Oitmaa}
\affiliation{School of Physics, University of New South Wales, Sydney 2052, Australia}

\begin{abstract}
We revisit the problem of a single hole moving in the
background of the two dimensional Heisenberg antiferromagnet.
The hole is loosely bound by an impurity potential.
We show that the bound state is generically a parity doublet:
there are parametrically close bound states of opposite parity.
Due to the degeneracy the bound state readily breaks local symmetries
of the square lattice and this leads to formation of the long range spiral 
distortion of the antiferromagnetic background. 
A direct analogy with van der Waals forces in atomic physics is discussed.
\end{abstract}

\date{\today}
\maketitle

\section{Introduction}
The problem of single hole binding by an attractive potential in the
 t-J model is of fundamental importance. In the physics of doped Mott 
insulatoris and in particular in the physics of the cuprate
superconductors, the system play a role analogous to the 
that played by the hydrogen atom in atomic physics.

We have in mind, for example, La$_2$CuO$_4$ with a La ion replaced by Sr.
Alternatively it may be Ca$_2$CuO$_2$Cl$_2$ with a Ca ion replaced by Na.
An important point is that the attractive center (Sr ion
in La$_2$CuO$_4$ or Na ion in Ca$_2$CuO$_2$Cl$_2$) sits in the center
of a square of four Cu sites. This means that the attractive potential
itself does not break the local square lattice symmetry.
There are various aspects of the bound state problem:  the symmetry/parity of 
the bound state, the structure of the spin fabric, and in the end
the particular value of the binding energy.
We argue that the symmetry issues are the most important ones.

There have been several studies of the bound state problem.
These are mainly small cluster exact 
dioganalizations~\cite{szcz,rabe,good,chen}.
A generic limitation of this approach is the small
cluster size and as a consequence sensitivity to boundary
conditions. In spite of this limitation  a very important observation 
has been made already in the early work~\cite{rabe}:
the ground state is almost degenerate with another state that has
opposite parity.
Dependent on the parameters of the model, the ground state belongs either to 
the two
dimensional E-representation of the $C_{4v}$ symmetry group of
the Hamiltonian or to the $A_1$ representation~\cite{chen}.
However, there is always a very low-lying excitation of opposite parity.

\begin{figure}[ht]
\vspace{10pt}
\includegraphics[width=0.45\textwidth,clip]{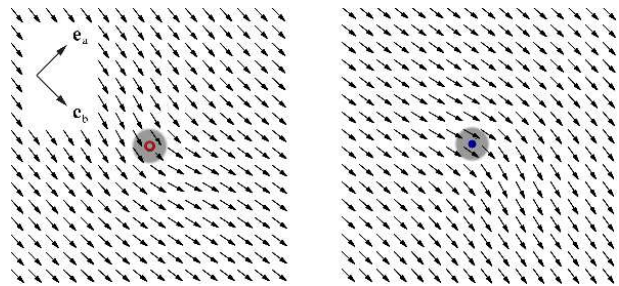}
\caption{\emph{(Color online). Distortion of the staggered spin fabric 
(small arrows) by
the Sr-hole bound state. The left picture corresponds to the pseudospin
directed out of the page and the right picture corresponds to the pseudospin
directed in the page. Shaded area corresponds to the hole localization
region. At large distances spins are directed along the orthorhombic
b-axis due to pinning by Dzyaloshinksi-Moriya and XY anisotropies.
}}
\label{H}
\end{figure}

A semiclassical solution of the bound state problem was obtained  in 
Ref.~\cite{sushkov05}
Generally a semiclassical approach can be justified in the limit of a 
large radius of the bound state.
According to the semiclassical solution the bound state generates  a long range 
($\propto  1/r$) 
spiral distortion of the spin fabric as it is shown in Fig.\ref{H}.
The figure shows staggered spins.
It is obvious from Fig.~\ref{H} that the semiclassical solution
does not have a definite parity. It does not fit in any
representation of the symmetry group of the square lattice.
So the solution spontaneously breaks the local square lattice symmetry.
On the other hand it is impossible to have a spontaneous violation of an exact
symmetry of the Hamiltonian in a finite system. Therefore, the exact parameter
that justifies the semiclassical solution~\cite{sushkov05} has remained unclear.
The purpose of the present work is to elucidate this parameter
and hence to elucidate the physical meaning of the solution with
violation of exact symmetries.
We show that the physics of this system is similar to the physics
of a hydrogen atom in an external electric field.
The parameter that justifies the semiclassical solution is the
small splitting between states of opposite parity.
The splitting scales as the binding energy squared and hence it is
infinitesimally small for  a shallow bound state.

The structure of the paper is as follows. In Section II we explain the
analogy with the hydrogen atom.
In particular, we consider the conditions when a long range tail of the
dipole electric field can be generated by the atom.
In Section III we briefly review known properties of an unbound
single hole moving in the antiferromagnetic background of the
$t-J$ model.
Section IV addresses the limiting case of very strong binding.
Here we present results of exact diagonalizations for the 4$\times$4
cluster embedded in an antiferromagnetic background.
In Sections V and VI we consider the weak binding limit and discuss
symmetry properties of the bound states.
Section VII addresses the parity breaking and  generation of the local
spin spiral. 
In Section VIII we exclude a possibility of the local Charge Density Wave 
(CDW) formation.
Our conclusions are presented in Section IX.

\section{Hydrogen atom}
Consider a hydrogen atom in the ground 1s state.
The size is about one Bohr radius, $a_B$.
Since the atom is neutral the electric field at distances $r \gg a_B$
decays exponentially.
Let us consider the same atom in the $n=2$ state,
either the positive parity 2s-state or negative parity 2p-state.
Importantly, they are degenerate.
Because of the degeneracy an infinitesimally small {\it external} electric field,
$E_{ext}\to 0$, will mix the opposite parity states
\begin{equation}
\label{2sp}
\psi=\frac{1}{\sqrt{2}}|2s\rangle+\frac{1}{\sqrt{2}}|2p_0\rangle \ .
\end{equation}
Here $|2p_0\rangle$ is the state with zero projection of the angular
momentum in the direction of the external electric field.
The state (\ref{2sp}) possesses a static electric dipole moment $d \sim e a_B$.
Hence a static dipole electric potential and a static 
electric field  are induced outside the atom, $r \gg a_B$
\begin{equation}
\label{fe}
\varphi_{ind}({\bf r})=-\frac{({\bm d}\cdot{\bm r})}{r^3} \ ,
\ \ \ 
{\bf E}_{ind}({\bf r})=
-\frac{\bf d}{r^3}+
\frac{3({\bf d}\cdot{\bf r}){\bf r}}{r^5}\ .
\end{equation}
Due to the small but nonzero energy splitting $\Delta$ between 
2s- and 2p-states (Lamb shift), one needs to apply a small but finite external 
field, $E_{ext}> \Delta/d$, to create the mixed state (\ref{2sp})
and hence to induce the dipole field (\ref{fe}).
Importantly, the induced field (\ref{fe}) is much larger than $E_{ext}$.

One can also look at the problem from another point of view.
Consider two hydrogen atoms each  in the $n=2$ state.
The attractive potential between the atoms has two distinct regimes
depending on the distance $r$ between the atoms. 
The characteristic distance $r_{\Delta}$ is defined by the condition
$d^2/r^3_{\Delta} \sim \Delta$.
If $a_B \ll r \ll r_{\Delta}$ the potential is
\begin{equation}
\label{Vat}
V\sim -\frac{d^2}{r^3} \ .
\end{equation}
In this regime the electric dipole fields of the two atoms lock to each other.
At $r\gg r_{\Delta}$  the interaction scales as $1/r^6$,
this is the usual van der Waals regime that is due to fluctuating dipoles.
We will argue below that the dipole distortion of the spin fabric
in the two-dimensional (2D) t-J model shown in Fig.~\ref{H}
is fully analogous to Eqs. (\ref{fe}),(\ref{Vat}).
The power in the 2D case is different, $\frac{1}{r^3} \to \frac{1}{r^2}$.
A more important difference is that in the t-J model the ground 
state itself is a parity doublet. The splitting in the doublet is 
parametrically small at small binding energy, it scales as 
$\Delta \propto \epsilon^2$, where $\epsilon$ is the binding energy.

\section{A free hole propagation in the $t-J$ model}

The 2D $t-J$ model was suggested two decades ago to describe the essential
low-energy physics of high-$T_{c}$ cuprates~\cite{PWA,Em,ZR}. In its
extended version, this model includes additional hopping matrix elements $%
t^{\prime }$ and $t^{\prime \prime }$ to 2nd and 3rd-nearest Cu
neighbors. The Hamiltonian of the $t-t^{\prime }-t^{\prime \prime }-J$ model
on the square Cu lattice has the form:%
\begin{eqnarray}
 \label{HtJ}
H_{t-J} &=&-t\sum_{\langle ij\rangle \sigma }c_{i\sigma }^{\dag }c_{j\sigma
}-t^{\prime }\sum_{\langle ij^{\prime }\rangle \sigma }c_{i\sigma }^{\dag
}c_{j^{\prime }\sigma }-t^{\prime \prime }\sum_{\langle ij^{\prime \prime
}\rangle \sigma }c_{i\sigma }^{\dag }c_{j^{\prime \prime }\sigma }  \notag \\
&+&J\sum_{\langle ij\rangle \sigma }\left( \mathbf{S}_{i}\mathbf{S}_{j}-{%
\frac{1}{4}}N_{i}N_{j}\right) . 
\end{eqnarray}%
Here, $c_{i\sigma }^{\dag }$ is the creation operator for an electron with
spin $\sigma $ $(\sigma =\uparrow ,\downarrow )$ at site $i$ of the square
lattice, $\langle ij\rangle $ indicates 1st-, $\langle ij^{\prime }\rangle $
2nd-, and $\langle ij^{\prime \prime }\rangle $ 3rd-nearest neighbor sites. 
The spin operator is $\mathbf{S}_{i}={\frac{1}{2}}c_{i\alpha }^{\dag }%
\mathbf{\sigma }_{\alpha \beta }c_{i\beta }$, and $N_{i}=\sum_{\sigma
}c_{i\sigma }^{\dag }c_{i\sigma }$ is
the number density operator. In addition to the Hamiltonian (\ref{H}) there
is the constraint of no double occupancy, which accounts for strong electron
correlations. 

The values of the parameters of the Hamiltonian (\ref{HtJ}) for
cuprates are known from neutron scattering, Raman 
spectroscopy, and ab-initio calculations.
For La$_2$CuO$_4$ the values are~\cite{tokura90,keimer92,andersen95}:
\begin{eqnarray}
\label{Jtt}
J &\approx& 140\,\text{meV}\to 1\ ,\nonumber\\
t&\approx& 450\,\text{meV}\ ,  \nonumber\\
t^{\prime } &\approx& -70\,\text{meV}\ , \nonumber\\
t^{\prime \prime }&\approx 35&\,\text{meV}\ .
\end{eqnarray}
Hereafter we set $J=1$, hence we measure energies in units of $J$.
In the present work we study generic properties of the extended
$t-J$ model. Therefore we will vary parameters $t$, $t^{\prime}$,
and $t^{\prime\prime}$ in a broad range.

 At zero doping (no holes), the $t$-$J$ model is equivalent to the Heisenberg 
model and describes the Mott insulator 
La$_2$CuO$_4$. The removal of a single electron from this Mott insulator, or in other words the injection of a 
hole, allows the charge carrier to propagate.

\begin{figure}[ht]
\vspace{10pt}
\includegraphics[width=0.25\textwidth,clip]{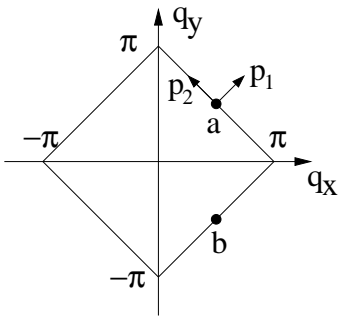}
\caption{\emph{Magnetic Brillouin Zone with a- and b-minima
of the hole dispersion}}
\label{valley}
\end{figure}

The properties of a free  single hole  in the $t$-$J$ model
are very well studied numerically: see Ref.~\cite{dagotto94} for a review. 
At values of parameters corresponding to the cuprates the
 dispersion of the hole dressed by magnetic 
quantum fluctuations has minima at the `nodal points'
 $\mathbf{q}_{0}=(\pm \pi /2,\pm \pi /2)$ see Fig.~\ref{valley}.
The typical value of the quasiparticle residue at these points is $Z\approx 0.3$.
By changing the sign of $t^{\prime}$ and $t^{\prime\prime}$ one can shift the 
dispersion minima to the `antinodal points' $(\pm\pi,0)$ and $(\pm\pi,0)$.
This situation corresponds to the electron doped cuprates.
We will argue below that the properties of the shallow bound state
are most interesting and rich in the regime when the minima of the
dispersion are at the nodal points. 
This is the regime which we consider in the present work. 
The dispersion of the hole dressed by magnetic quantum fluctuations
is quadratic in the vicinity of $\mathbf{q}_{0}$,
\begin{eqnarray}
\label{eq}
\epsilon \left( \mathbf{p}\right) \approx 
\frac{1}{2}\beta_1 p_1^{2}
+
\frac{1}{2}\beta_2 p_2^{2}
\ ,
\end{eqnarray}
where ${\bf p}={\bf q}-\mathbf{q}_{0}$.
We set the lattice spacing to unity, 3.81\thinspace \AA $\,\rightarrow $
\thinspace 1. In Eq.(\ref{eq}) $p_1$ is directed along the nodal direction
and $p_2$ is directed along the face of the Magnetic Brillouin Zone (MBZ),
see Fig.~\ref{valley}.
At values of the hopping parameters presented in Eq. (\ref{Jtt})
the inverse masses are~\cite{sushkov97}, $\beta_1\approx\beta_2\approx 2.5$.

It is instructive  to consider also the weak coupling limit $t \ll J$,
$t^{\prime}=t^{\prime\prime}=0$. In this limit
the quasiparticle residue is close to unity, $Z=1-O(t^2/J^2)$, while the 
dispersion reads~\cite{hamer98}
\begin{eqnarray}
\label{std}
\epsilon_{\bm q}&=&4t^{\prime}_{eff}\cos q_x\cos q_y
+2t^{\prime\prime}_{eff}\left(\cos 2q_x +\cos 2q_y\right)\ ,\nonumber\\
t^{\prime}_{eff}&\approx&0.25\frac{t^2}{J} \ , \nonumber\\
t^{\prime\prime}_{eff}&\approx&0.28\frac{t^2}{J} \ , \nonumber\\
\beta_1&=&4t^{\prime}_{eff}+8t^{\prime\prime}_{eff}\approx 3.26\frac{t^2}{J}
 \ , \nonumber\\
\beta_2&=&-4t^{\prime}_{eff}+8t^{\prime\prime}_{eff}\approx 1.23\frac{t^2}{J} \ .
\end{eqnarray}

\section{Hole binding in the strong coupling limit}
\begin{figure}[ht]
\vspace{10pt}
\includegraphics[width=0.25\textwidth,clip]{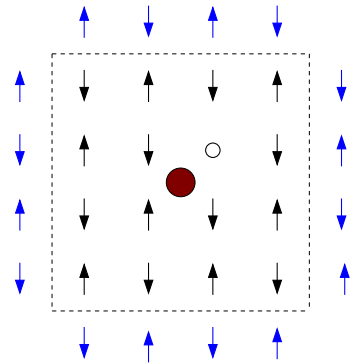}
\caption{\emph{(Color online). The strong binding limit.
Attraction to the potential center (a filled red circle in
the center of the plaquette) is so strong that the hole
(small empty circle) can hop only within the few sites
around the attractive center.
In the exact  cluster diagonalization we assume that
the perimeter spins (blue arrows outside of the dashed square) 
are static with $\langle S_z\rangle=\pm 0.3$. 
}}
\label{sB}
\end{figure}

We include a site-dependant potential attraction to an impurity, 
\begin{eqnarray}
\label{HU}
H&=&H_{t-J}+H_U \ , \nonumber\\
H_U&=&=
\sum_{i \sigma }U_ic_{i\sigma }^{\dag }c_{i\sigma} \ .
\end{eqnarray}
A very important point is that the potential $U_i$ is symmetric around the
center
of a plaquette, see Fig.~\ref{sB}. Note that the Hamiltonian (\ref{HU}) 
is written in terms of
electrons. Repulsion for electrons $U_i > 0$ corresponds to hole
attraction.
Concerning the potential we will consider two possibilities.
The first possibility is the short range potential which is
nonzero only at four nearest sites, see Fig.~\ref{sB},
\begin{equation}
\label{Usr}
U_i=U\sum_j\delta_{ij}\ ,
\end{equation}
where $j$ runs over four nearest sites.
The second possibility is the long range Coulomb interaction
\begin{equation}
\label{UC}
U_i=\frac{Q}{\sqrt{r_i^2+1}}\ ,
\end{equation}
where $Q$ is a dimensionless charge which sits at a distance 
one lattice spacing above the plane.

Let us consider first the local potential (\ref{Usr}) in the
strong coupling limit, $U \gg t$.
The solution of the bound state problem in this limit is qualitatively clear.
There are degenerate states with $S_z=\pm\frac{1}{2}$.
At each value of $S_z$ there are states of positive and negative parity.
Let us consider the lowest bound state in each parity sector.
Note that there is only one lowest state in each sector.
This is contrary to the common wisdom that the negative parity
states are doubly degenerate due to the symmetry of the square lattice
(E-representation of C$_{4v}$).
The point is that we consider hole binding on the spin background with
spontaneously broken SU(2) symmetry. In combination  with the impurity potential
this  breaks the symmetry of the square lattice and hence destroys
degeneracy of the negative parity states.
Binding  energies of the lowest bound states are
\begin{eqnarray}
\label{ub}
\epsilon_{\pm}\approx -U\pm\frac{\Delta}{2} \ .
\end{eqnarray}
Here the sign $\pm$ denotes the parity of the bound state.
We define the binding energy $\epsilon$ in the standard way:
this is the energy of the bound state taken with respect to the minimum
energy of a free hole. So $\epsilon$ is always negative.
To find the parity splitting  $\Delta$ in the strong coupling  limit we have 
performed  exact diagonalizations of the $t-t'-J$ model
on the 16 site cluster shown in Fig.~\ref{sB}.
We have already pointed out that it is qualitatively important to perform 
the diagonalization on the state with spontaneously broken SU(2) symmetry.
Therefore we put the cluster in the environment of static perimeter 
spins shown in Fig.~\ref{sB} by blue arrows outside of the dashed
square. The magnetization of each static
spin has the Heisenberg model value, $\langle S_z\rangle=\pm 0.3$.
\begin{table}[ht]
\begin{tabular}{|lr|cc|cc|cc|}
&&t'=0& &t'=0.5& & t'=-0.5&\\
\hline
t& U &$\Delta$\ & \ $r_{rms}$&$\Delta$\ & \ $r_{rms}$&$\Delta$\ & \ $r_{rms}$  \\
&  & & & & & &\\ 
$ 0.25$ &  0 & {\large 0.002} \ \ & \ \ {\large 2.05} & 0.562 \ \ & \ \ 1.34 &-0.558 \ \ & \ \ 1.33 \\  
$ 0.25$ & 10 & {\large 0.024} \ \ & \ \ {\large 0.71} & 0.570 \ \ & \ \ 0.72 &-0.577 \ \ & \ \ 0.72\\  
 & & & && && \\ 
$ 0.5$ &  0  & {\large 0.037} \ \ & \ \ {\large 1.83} & 0.555 \ \ & \ \ 1.33 &-0.416 \ \ & \ \ 1.30\\  
$0.5$&     10& {\large 0.075} \ \ & \ \ {\large 0.71} & 0.567 \ \ & \ \ 0.72 &-0.547 \ \ & \ \ 0.72\\  
 & & &&& &&  \\ 
$ 1.0$ &   0 & {\large 0.096} \ \ & \ \ {\large 1.60} & 0.503 \ \ & \ \ 1.32 &{\large -0.079} \ \ & \ \ {\large 1.09}\\  
$1.0$  &  10 & {\large 0.172} \ \ & \ \ {\large 0.72} & 0.558 \ \ & \ \ 0.73 &{\large -0.283} \ \ & \ \ {\large 0.73}\\  
 & & &&& &&  \\ 
$ 2.0$ &   0 & {\large 0.062} \ \ & \ \ {\large 1.26} & 0.424 \ \ & \ \ 1.25 &{\large  0.010} \ \ & \ \ {\large 1.25}\\  
$2.0 $ &  10 & {\large 0.253} \ \ & \ \ {\large 0.76} & 0.503 \ \ & \ \ 0.76 &{\large -0.040} \ \ & \ \ {\large 0.76}\\  
 & & &&& &&  \\ 
$ 3.0$ &   0 & {\large 0.067} \ \ & \ \ {\large 1.18}  & 0.337 \ \ & \ \ 1.20 &{\large  0.009} \ \ & \ \ {\large 1.21}\\  
$ 3.0$ &  10 & {\large 0.237} \ \ & \ \ {\large 0.79}  & 0.431 \ \ & \ \ 0.79 &{\large  0.004} \ \ & \ \ {\large 0.80}\\  
 & & &&& &&  \\ 
$ 4.0$ &   0 & {\large 0.066} \ \ & \ \ {\large 1.17}  & 0.280 \ \ & \ \ 1.18 &{\large  0.004} \ \ & \ \ {\large 1.19}\\  
$4.0$ &   10 & {\large 0.181} \ \ & \ \ {\large 0.82}  & 0.360 \ \ & \ \ 0.82 &{\large -0.031} \ \ & \ \ {\large 0.82}\\ 
\hline
\end{tabular}\\
\caption{\it Exact diagonalization of the 16-site cluster (Fig.~\ref{sB}).
The ground state parity doublet energy splitting and the rms charge 
radius of the ground state for several values of $t$ and $t'$ and for two values
of the confining potential $U$. According to Eq.(\ref{ub})
$\Delta >0 $ corresponds to the negative parity of the ground state
and $\Delta < 0 $ corresponds to the positive parity of the ground state.
The bound state results for values of $t$ and $t'$ that
correspond to the free hole dispersion minima at the nodal points, 
$(\pm \pi/2,\pm\pi/2)$, are presented by {\large large} font.
}
\end{table}
Values of the splitting $\Delta$ within the parity doublet obtained 
by the 16 site cluster exact diagonaization are presented in Table I.
In the same table we present values of the rms charge radius of the lowest 
bound state.
The results are presented for several values of $t$ and $t'$.
The value of $t''$ in this calculation is zero: the cluster is too small
to account for long range hopping.
We have performed the calculation for two values of the confining potential,
U=0 and U=10. In an infinite system the case U=0 certainly does not correspond
to any binding. However, for the cluster, due to the imposed boundary conditions, the
case U=0 describes a well localized state of the hole;
in this sense it is bound.
Our numerical results qualitatively agree with those of previous 
publications~\cite{szcz,rabe,good,chen}.
A detailed quantitative comparison is not possible because the previous publications
have considered spin symmetric clusters while we impose a spontaneous
violation of the SU(2) symmetry via boundary conditions. 

The dispersion of a free hole for various values of $t$ and $t'$ is well
known from previous work~\cite{dagotto94,sushkov97,hamer98}.
The bound state results in Table I for values of $t$ and $t'$ that
correspond to the free hole dispersion minima at the nodal points, $(\pm \pi/2,\pm\pi/2)$,
are presented by {\large large} font.
For all other values of $t$ and $t'$ the dispersion minima are at the antinodal
points, $(\pm \pi,0)$, $(0,\pm\pi)$, or at the $\Gamma$-point, ${\bf k}=0$. 
The corresponding bound state results in Table I are shown by standard font.
Results presented in Table I lead to the following observations.\\
1) Values of the parity splittings for the `nodal cases' (large font)
 are very small 
compared to typical scales in the problem. This conclusion is in agreement
with previous observation~\cite{rabe}.\\
2) Values of the splittings for other cases (`antinodal' and `$\Gamma$-point') 
are substantially larger.\\
3) For each particular `nodal' set of $t$ and $t'$ the splitting $\Delta$ 
for U=0 is systematically smaller than that for U=10.

The first and second observations
indicate that the bound state is close to the parity degeneracy
in the case of the nodal minima of the free hole dispersion.
This is why in the present work we
concentrate  on this case.
According to the third observation the parity splitting is rapidly decreasing 
when the radius of the bound state increases. In following section
we consider shallow bound states of large radius to confirm these
conclusions.

\section{Hole binding in the weak coupling limit: the leading approximation}
Let us look at the binding problem in the weak coupling limit, $U \to 0$,
$\epsilon \to 0$.
In this case the shallow bound state can be built with a hole either from 
the a- or the b-valley of the dispersion, see Fig.~\ref{valley}.
Hence the bound state has the valley index and the corresponding
wave function reads
\begin{equation}
\label{psis}
\psi_{\alpha\sigma}({\bm r})=e^{i{\bm q}_{\alpha}\cdot{\bm r}_{\sigma}}
\chi_{\alpha}(r_{\sigma}) \ ,
\end{equation}
where $\alpha=a,b$ shows the valley and
$\sigma=\uparrow,\downarrow$ shows the magnetic sublattice 
along which the hole is propagating, $r_{\sigma}$ is position on this
sublattice.
Note that here we have in mind a real propagation.
There are also virtual hoppings of the hole to the opposite sublattice.
These virtual hoppings lead to formation of the free hole dispersion that
was
discussed in Section III.
The z-projection of the hole spin is $S_z=-\sigma$ .
The oscillating exponential dependence  $e^{i{\bm q}_{\alpha}\cdot{\bm r}_{\sigma}}$
in (\ref{psis})  is due to the momentum 
 ${\bm q}_{\alpha}=(\pm \pi /2,\pm \pi /2)$ 
that corresponds to the valley minimum.
The very smooth function $\chi(r)$ exponentially decaying at infinity
is due to the hole binding to the potential.
In the case of the  Coulomb field, Eq.(\ref{UC}), the wave function is~\cite{com1}
\begin{eqnarray}
\label{chiC}
\chi&=&\sqrt{\frac{2}{\pi}}\kappa
e^{-\sqrt{2|\epsilon|(r_1^2/\beta_1+r_2^2/\beta_2)}}\ , \nonumber\\
\kappa&=&\sqrt{\frac{2|\epsilon|}{\sqrt{\beta_1\beta_2}}} \ .
\end{eqnarray}
The components $r_1$ and $r_2$ in Eq.(\ref{chiC}) are projections
of ${\bf r}$ on directions 1 and 2 corresponding to the particular
valley, see Fig.~\ref{valley}.
In the case of the  local attraction Eq.(\ref{Usr}), the wave function is
\begin{eqnarray}
\label{chiL}
\chi&=&\frac{\kappa}{\sqrt{\pi}}K_0(\sqrt{2|\epsilon|(r_1^2/\beta_1+r_2^2/\beta_2)})
\ , 
\end{eqnarray}
where $K_0$ is the Bessel function of the second kind.
Note that for both Eqs.(\ref{chiC}) and (\ref{chiL}) the root mean square
radius of the bound state scales as
\begin{equation}
\label{rrm}
r_{rms}\propto \frac{1}{\sqrt{|\epsilon|}} \ .
\end{equation}
It is easy to see that a change of sign of ${\bm q}_{\alpha}$ leads only
to a common phase factor in the wave function (\ref{psis}),
so this is the same wave function.
There are only two distinct possibilities:
the a-minimum, ${\bm q}_{a}=(\pi /2, \pi /2)$, and the b-minimum,
${\bm q}_{b}=(\pi /2,-\pi /2)$.
Thus, there are two degenerate quantum states for each value of $S_z$.

We put the potential center at the origin of the coordinate system.
Then, according to Eq. (\ref{psis}) and Fig.~\ref{sB} 
(in this case one has to remove the cluster boundary and extend the figure
up to infinity) the $|\uparrow\rangle$
wave functions read
\begin{eqnarray}
\label{psis1}
&&\psi_{a\uparrow}({\bm r})=-ie^{i\frac{\pi}{2}x_{\uparrow}+i\frac{\pi}{2}y_{\uparrow}}
\chi_a(r) \ ,\nonumber\\
&&\psi_{b\uparrow}({\bm r})=e^{i\frac{\pi}{2}x_{\uparrow}-i\frac{\pi}{2}y_{\uparrow}}
\chi_b(r) \ ,\nonumber\\
&&x_{\uparrow}=\frac{1}{2}+m \ , \nonumber\\
&&y_{\uparrow}=\frac{1}{2}+n \ ,
\end{eqnarray}
where  both $m+n$ and $m-n$ are integer and even.
Similarly the $|\downarrow\rangle$ wave functions are
\begin{eqnarray}
\label{psis2}
&&\psi_{a\downarrow}({\bm r})=e^{i\frac{\pi}{2}x_{\downarrow}+i\frac{\pi}{2}y_{\downarrow}}
\chi_a(r) \ ,\nonumber\\
&&\psi_{b\downarrow}({\bm r})=
-ie^{i\frac{\pi}{2}x_{\downarrow}-i\frac{\pi}{2}y_{\downarrow}}
\chi_b(r) \ ,\nonumber\\
&&x_{\downarrow}=\frac{1}{2}+m \ , \nonumber\\
&&y_{\downarrow}=-\frac{1}{2}+n \ .
\end{eqnarray}
We put an additional factor $-i$ in $\psi_{a\uparrow}$ and 
$\psi_{b\downarrow}$ to make these wave functions real.
Under the parity operation $x \to -x$ and $y \to -y$, the function
$\chi(r)$ does not change.
Therefore parities of states (\ref{psis1}) and (\ref{psis2})
are determined by the phase factors, and the parities are
\begin{eqnarray}
\label{P}
P_{a\uparrow}&=&e^{i\pi x_{\uparrow}+i\pi y_{\uparrow}}=e^{i\pi}=-1 \ ,\nonumber\\
P_{b\uparrow}&=&e^{i\pi x_{\uparrow}-i\pi y_{\uparrow}}=e^{i0}=+1 \ ,\nonumber\\
P_{a\downarrow}&=&e^{i\pi x_{\downarrow}+i\pi y_{\downarrow}}=e^{i0}=+1 \ ,\nonumber\\
P_{b\downarrow}&=&e^{i\pi x_{\downarrow}+i\pi y_{\downarrow}}=e^{i\pi}=-1 \ .
\end{eqnarray}
Thus, in the leading weak coupling limit approximation, $\epsilon \to 0$,
the ground state is a degenerate parity doublet for each value of $S_z$.
This explains why values of the parity splitting presented in Table I
by large font are very small. The next Section is addressed to the mechanism 
that lifts the exact parity degeneracy.
\\

\section{Hole binding in the weak coupling limit: the subleading 
approximation}
In the present section we  demonstrate that in the subleading
weak binding approximation, $\epsilon \to 0$, the parity degeneracy of 
the ground state obtained in the previous section is lifted:
the parity splitting scales as $\Delta \propto \epsilon^2$.
Because of this scaling the splitting is very small compared to the 
binding energy, $\Delta \ll \epsilon$.

\begin{figure}[ht]
\vspace{10pt}
\includegraphics[width=0.45 \textwidth,clip]{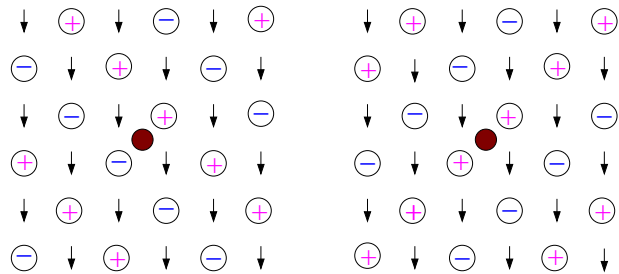}
\caption{\emph{(Color online). 
Left: signs of the wave function $\psi_{a\uparrow}$,
Right: signs of the wave function $\psi_{b\uparrow}$.
The potential center is shown by the filled red circle.
}}
\label{WF}
\end{figure}

Signs of the hole wave functions $\psi_{a\uparrow}$ and $\psi_{b\uparrow}$
given by Eqs.~(\ref{psis1}) are shown in Fig.~\ref{WF}.
In this case the hole is moving on the `up' sublattice.
To avoid misunderstanding we stress that there is only one hole,
we do not show spins up in Fig.~\ref{WF} just to make the figure less 
busy.
The figure clearly demonstrates that the states have opposite parities
and different diagonal momenta, ${\bm q}_a=(\pi/2,\pi/2)$ and               
${\bm q}_b=(\pi/2,-\pi/2)$.

It is clear from Fig.~\ref{WF} that the difference in energy between states 
$\psi_{a\uparrow}$ and $\psi_{b\uparrow}$
arises due to diagonal hopping  of the hole in the vicinity of the potential,
$\Delta \propto t'_{eff}|\chi(0)|^2$, where $t'_{eff}$ is the effective diagonal
hopping that is due to the bare $t'$ and also due to higher orders in $t$,
see e.g. Eq.(\ref{std}). Moreover, the splitting cannot be just proportional
to $|\chi(0)|^2$; the splitting must contain a gradient of $\chi$ because there 
is no a splitting for free hole propagation when $\chi=const$.
The first power of the gradient in the energy splitting is forbidden by parity.
Thus we come to the following formula for the energy splitting
\begin{equation}
\label{dd2}
\Delta \propto t'_{eff}|\nabla\chi(0)|^2 \ .
\end{equation}
The formula contains the second power of gradient, so it is allowed by parity.
Having in mind Eqs.(\ref{chiC}), (\ref{chiL}) and using Eq.(\ref{dd2}) we conclude
that
\begin{equation}
\label{dd3}
\Delta \propto t'_{eff}\kappa^4 \propto t'_{eff}\epsilon^2\propto
\frac{t'_{eff}}{r_{rms}^4} \ .
\end{equation}
We stress that this formula follows from general symmetry considerations
based on degeneracy of the free hole dispersion at the four
nodal points ${\bm q}=(\pm\pi/2,\pm\pi/2)$.
The symmetry arguments  certainly do not allow to determine a coefficient in
Eq. (\ref{dd3}). However, they do allow us to determine the scaling law given
by (\ref{dd3}).

\begin{figure}[ht]
\vspace{10pt}
\includegraphics[width=0.5 \textwidth,clip]{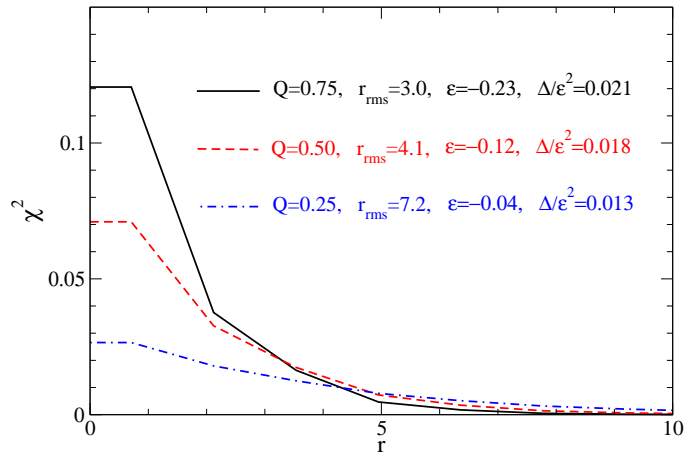}
\caption{\emph{(Color online). 
The shallow Coulomb bound  state wave function squared, $\chi^2$,
versus radius.
Wave functions are shown for for three values of the dimensionless
charge: $Q=0.75$, $Q=0.5$, and $Q=0.25$.
Values of the effective hopping parameters are $t'_{eff}=0.1$, $t''_{eff}=0.25.$
In the legend, for every value of $Q$ we also present the rms radius
of the bound state, $r_{rms}$; the binding energy, $\epsilon$; and the ratio of 
the parity doublet splitting over the binding energy squared, $\Delta/\epsilon^2$.
}}
\label{chi2}
\end{figure}

It is helpful to support the general considerations presented in the previous 
paragraph by a numerical calculation. Such a calculation in the regime $t > J$
is hardly possible. However, in the regime $t < J$ the calculation can be
performed using results of Ref.~\cite{hamer98} summarized in Eq.(\ref{std}).
According to the results the spin quantum fluctuations can be integrated out
and the hole propagation on the sublattice up is described by the following 
effective Hamiltonian
\begin{equation}
\label{heff}
H_{eff}=t^{\prime }_{eff}\sum_{\langle ij^{\prime }\rangle}
h_{i}^{\dag}h_{j^{\prime }}+t^{\prime \prime }_{eff}\sum_{\langle ij^{\prime \prime
}\rangle}h_{i}^{\dag }h_{j^{\prime \prime }} \ .
\end{equation}
Here $h_{i}^{\dag}$ is the holon creation operator on the site $i$; all the
sites $i$, $j^{\prime }$ and $j^{\prime \prime }$ belong to the sublattice up.
To be specific we consider here the Coulomb attraction (\ref{HU}), (\ref{UC}).
The attractive interaction written in terms of holon operators reads
\begin{equation}
\label{HCh}
H_C=-\sum_iU_ih_{i}^{\dag}h_{i}\ ,
\end{equation}
where $U_i$ is given by Eq.(\ref{UC}).
The Hamiltonian $H_{eff}+H_C$ can be easily diagonalized numerically on a very 
large cluster. Results of diagonalizations for 30$\times$30 cluster with
$t'_{eff}=0.1$, $t''_{eff}=0.25$ and for three values of the dimensionless charge
$Q=0.75$, $Q=0.5$, and $Q=0.25$ are shown in Fig.~\ref{chi2}.
In this figure we show the holon probability distribution for shallow bound states,
and in the legends we present values of the rms radius, $r_{rms}$;
the binding energy, $\epsilon$; and the ratio $\Delta/\epsilon^2$.
According to the data in Fig.~\ref{chi2}, in the limit $\epsilon \to 0$ 
the parity splitting $\Delta$  is decaying even slightly faster than 
$\propto \epsilon^2$.
Most likely the small deviation from the expected $\epsilon^2$ scaling is due to 
the finite cluster size. 
We have also checked that the splitting $\Delta$ vanishes at $t'_{eff}=0$.
Altogether the numerical results presented in Fig.~\ref{chi2} confirm the scaling
law given by Eq.(\ref{dd3}).

The conclusion of the present section is that the ground state parity splitting
is decaying $\Delta \propto \epsilon^2 \propto 1/r_{rms}^4$ when the binding 
energy  is  decreasing, $\epsilon \to 0$, $r_{rms}\to \infty$. 
To estimate the coefficient in this dependence at $t >J$ one can refer to results 
of exact numerical diagonalizations presented in Table I.
It is known experimentally that in very lightly doped La$_{2-x}$Sr$_x$CuO$_4$
a hole binding energy to Sr ion is about $\epsilon \approx -10$ meV,
the bound state `wave vector' is $\kappa\approx 0.4$, and the rms 
radius of the bound state is $r_{rms}\approx 3$, as discussed in
Ref.~\cite{sushkov05} Estimates based on results derived in 
Sections IV and VI show that the expected parity splitting of the ground 
state in this case is a small fraction of 1 meV.
Therefore, parity breaking is practically a zero mode of the system.

\section{Parity breaking and formation of the local spin spiral}
According to the discussion in previous sections a single hole bound state
in the t-J model always has a definite value of the spin projection on the 
direction of staggered magnetization, $S_z=\pm 1/2$, and it always has a 
definite parity. Dependent on parameters, $t$, $t'$, etc, the 
ground state parity can be 
positive or negative, but it is definite.
There is no local spin spiral at this stage.
Very close to the ground state there is always a state of opposite parity.
Wave functions of these states are given by Eqs. (\ref{psis1}) and (\ref{psis2}),
and parities are given by Eq. (\ref{P}).
Now, following the 2s-2p hydrogen atom scenario discussed in Section II,
we can mix the opposite parity states by a weak  external perturbation.
There are two possibilities:
1)mixing of states with different $S_z$ that belong to the same hole pocket,
2)mixing of states with the same $S_z$ that belong to different hole pockets.
In the present section we consider the first possibility that leads to formation
of a local spin spiral shown in Fig.~\ref{H}.
The second possibility could lead to formation of a CDW. However, we show in the
following section that this possibility is energetically unfavorable.

\begin{figure}[ht]
\vspace{10pt}
\includegraphics[width=0.25\textwidth,clip]{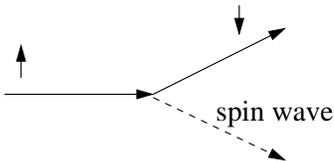}
\caption{\emph{Hole-spin-wave vertex corresponding to the Hamiltonian (\ref{hi})
}}
\label{vert}
\end{figure}

Thus, let us mix the opposite parity states with different $S_z$ that belong to 
the same pocket.
To do so, we impose a very weak uniform spin twist on the system.
At this stage it becomes convenient to use the notation of the non-linear $\sigma$-model.
In this notation the unit vector ${\vec n}({\bm r})$ shows direction of staggered
spins. In the antiferromagnetic state the spins are directed along the z-axis in the
spin space, ${\vec n}=n_z=(0,0,1)$. The uniform spin twist means that the spin
direction ${\vec n}$ rotates around a unit vector ${\vec\xi}$ that is orthogonal to
the z-axis. So locally we can write
\begin{equation}
\label{rw}
\delta{\vec n}({\bm r})=({\bm Q}\cdot{\bm r})[{\vec \xi}\times {\vec n}] \ .
\end{equation}
Here ${\bm Q} \ll 1 $ is the wave vector of the imposed twist.
 Let us direct ${\bm Q}$ along the b nodal direction, see Fig.~\ref{valley},
${\bm Q}=Q{\bm e}_b$, where ${\bm e}_b$ is the b-nodal unit vector.
It is worth noting that generally directions in spin space and directions in
the coordinate space are completely independent.
The interaction of a hole with the deformation of the spin fabric is of the following 
form~\cite{shraiman88}
\begin{equation}
\label{hi}
H_{int}=-\sqrt{2}g{\vec \sigma}\times[{\vec n}\times({\bm e}\cdot{\bm \nabla}{\vec n}] \ ,
\end{equation}
where ${\bm e}$ is a nodal unit vector corresponding to the particular hole,
${\vec\sigma}$ is the Pauli matrix acting on pseudospin of the hole.
Note that in the notation of the original t-J model the effective Hamiltonian
(\ref{hi}) is just the usual hole-spin-wave vertex shown in Fig.~\ref{vert}.
Therefore  the coupling constant is $g\approx Zt\approx 1$,
where $Z\approx J/t$ is the quasiparticle residue of the hole.

Since ${\bm Q}=Q{\bm e}_b$, the interaction (\ref{hi}) does not mix states
$\psi_{a\uparrow}$ and $\psi_{a\downarrow}$ however, it does mix states
$\psi_{b\uparrow}$ and $\psi_{b\downarrow}$.
The corresponding interaction energy is
\begin{equation}
\label{eint}
\delta E_{int}=-\sqrt{2}g Q\langle \psi|({\vec \xi}\cdot{\vec\sigma}|\psi\rangle \ . 
\end{equation}
If the interaction energy is larger than the parity doublet splitting,
\begin{equation}
\label{QQ}
\sqrt{2}g Q > \Delta\ ,
\end{equation}
 the bound state wave function becomes a mixture of the opposite parity states
\begin{equation}
\label{cbb}
\psi=\frac{1}{\sqrt{2}}\left(\psi_{b\uparrow}+e^{i\alpha}\psi_{b\downarrow}\right)
\end{equation}
with the phase $\alpha$ determined by the condition 
$\langle \psi|{\vec \sigma}|\psi\rangle={\vec \xi}$.
Thus, the uniform spin twist Q is completely analogous to a weak uniform 
electric filed
$E_{ext}$ applied to hydrogen atom as has been discussed in Section II.
The wave function mixing (\ref{cbb}) is analogous to the mixing (\ref{2sp}).
Estimates based on values of $\Delta$ obtained in previous sections
show that for a bound state with radius $r_{rms}=3$ the twist $Q=0.001-0.002$
is already sufficient to break the parity according to Eqs. (\ref{QQ}),(\ref{cbb}).
Note that this small value of Q corresponds to a wavelength of about 5000 lattice spacing.

The state (\ref{cbb}) possesses a spin-flip dipole moment and hence it creates a 
long range distortion of the spin fabric as has been discussed in 
Ref.~\cite{sushkov05}
\begin{equation}
\label{dn}
\delta{\vec n}_{ind}=[{\vec \xi}\times{\vec n}]\frac{g}{\sqrt{2}\pi\rho_s}
\frac{({\bm e}\cdot{\bm r})}{r^2}\left[1-e^{-2\kappa r}(1+2\kappa r)\right] \ .
\end{equation}
Here $\rho_s\approx 0.18J$ is the spin stiffness of the Heisenberg model,
and $\kappa$ is the inverse radius of the charge core, see Eq.(\ref{chiC}).
Eq. (\ref{dn}) describes the local spiral depicted in Fig.\ref{H},
the local spin spiral is fully analogous to the long range scalar 
potential $\varphi_{ind}({\bm r})$
generated by an excited hydrogen atom in a tiny external electric field, see 
Eq. (\ref{fe}).

To derive Eq.(\ref{cbb}) and hence to justify the local spin spiral (\ref{dn}) we have
introduced a tiny external spin twist that enforces the parity breaking.
Alternatively, one can consider an interaction between two holes
bound to two impurities separated by a large distance $r$. Then there is no 
need for any external twist.
Spin spirals induced by different holes lock each other.
Hence the spin spiral induced hole-hole interaction is~\cite{luscher07}
\begin{equation}
\label{vv}
E_S \sim -\frac{g^2}{4\rho_s}\frac{1}{r^2} \ .
\end{equation}
This formula is valid at $r < r_{\Delta}$, while at $r > r_{\Delta}$ the interaction 
is $E_S \propto 1/r^4$.
Once more, this is absolutely similar to the case of two Hydrogen atoms, see 
Eq.(\ref{Vat}). Estimates based on values of $\Delta$ obtained in Sections 
IV,VI show that for  bound states with radius $r_{rms}=3$ the value of the 
crossover distance
is $r_{\Delta}\sim 50$. So practically Eq.(\ref{vv}) is always valid.
To restore dimension in (\ref{vv}) one has to recall that $g\approx J\approx 140$meV,
$\rho_s \approx 0.18J$ while dimensionless $r$ is  expressed in units of lattice spacing.

\section{Parity breaking and possible formation of the charge density wave}
We consider now a possible mixing of the bound states with the same $S_z$ that 
belong to different hole pockets.
Since the spin is not changed there is no deformation of the spin fabric,
and a usual electrostatic potential can mix the states.
However, the spatial wave functions from different pockets
differ by momentum ${\bm K}=(\pi,0)$, or ${\bm K}=(0,\pi)$.
Therefore, to generate the mixing the electrostatic potential must be modulated
at this momentum. 
So, the mechanism can produce a CDW with the wave vector ${\bm K}$.
Let $\phi_{\bm k}$ is a Fourier component of the external
electrostatic potential. The component interacts with the corresponding matrix
element of charge density
\begin{eqnarray}
\label{cd}
\rho_{\bm k}&=&\int\psi_{b\uparrow}^*({\bm r})e^{-i{\bm k}\cdot{\bm r}}\psi_{a\uparrow}d^2r
=\int\chi^2({\bm r})e^{i({\bm K- \bm k})\cdot{\bm r}}d^2r\nonumber\\
&=&\frac{8\kappa^3}{[4\kappa^2+({\bm K- \bm k})^2]^{3/2}} \ .
\end{eqnarray}
We have used here Eqs.(\ref{psis1}) and (\ref{chiC}).
We assume that $\beta_1=\beta_2$, this allows to evaluate the integral in (\ref{cd})
analytically. Numerical integration shows that Eq.(\ref{cd}) is approximately
valid even with non-equal inverse masses.
For example at $\beta_1/\beta_2=4$ the deviation from the analytical expression
Eq.(\ref{cd}) does not exceed a few per cent.

We proceed now directly to the Coulomb interaction between two holes bound to two
different impurities separated by large distance $r$.
Since the system is two-dimensional the electrostatic potential created
by a charge density component $\rho_{\bm k}$ is
\begin{equation}
\label{fi}
\phi_{\bm k}=\frac{2\pi}{k}\rho_{\bm k} \ .
\end{equation}
Therefore the Coulomb interaction energy between two spatially separated bound states
reads
\begin{equation}
\label{EC}
E_C=-\int\frac{2\pi}{k}\rho_{\bm k}^2e^{i{\bm k}\cdot{\bm r}}\frac{d^2k}{(2\pi)^2} \ .
\end{equation}
Note that the sign is negative because the system always tunes up the mixing phases
to reduce energy.
Evaluation of (\ref{EC}) with account of (\ref{cd}) is straightforward. 
In the  limit $\kappa r \gg 1$ the result reads
\begin{equation}
\label{EC1}
E_C=-\left(\frac{e^2}{a\epsilon_e}\right)\frac{\kappa^2}{2\pi}(2\kappa r)^2
\sqrt{\frac{\pi}{4\kappa r}}e^{-2\kappa r} \ .
\end{equation}
Here $\kappa$ is dimensionless and we put the factor 
$e^2/(a\epsilon_e)\approx 95meV$ 
to restore the dimension of energy ($e$ is the elementary charge, 
$a=3.81\thinspace \AA$ is the lattice spacing and $\epsilon_e \approx 40$ 
is the dielectric constant).

Now we can compare the CDW Coulomb interaction (\ref{EC1}) with the spin-spiral
interaction (\ref{vv}). In LSCO the `wave vector' of the bound state is
$\kappa \approx 0.4$, see Ref.~\cite{sushkov05} Let us take $r=4$ that corresponds
to the average distance between bound states at the doping level $x\approx 0.06$.
With these parameters one finds $E_C \lesssim 1$meV while $E_S \sim 15$meV.
Thus formation of the CDW is energetically unfavorable compared to formation
of the spin spiral .

\section{conclusions}
We have considered a single hole injected into a two dimensional Mott insulator
on a square lattice with a long range antiferromagnetic order.
The system is described by the extended t-J model.
An important point is that minima of the hole dispersion are
at nodal points $(\pm\pi/2,\pm/\pi/2)$.
The hole is bound by an impurity potential. The impurity is located at a center
of the lattice plaquette, so the potential itself does not break the local square 
lattice symmetry.

1) All bound states have definite parity and they are doubly degenerate with respect to 
the spin projection on the axis of the staggered magnetization, $S_z=\pm\frac{1}{2}$.

2)The ground state always has a very close state of opposite parity (parity doublet).
  For shallow bound states splitting within the parity doublet scales as 
  $\Delta \propto \epsilon^2$,  where $\epsilon$ is binding energy.

3)For shallow bound states the parity splitting $\Delta$ is extremely small.
  Therefore an extremely small external twist of the spin fabric breaks parity.
  The breaking creates a long range spiral distortion of the spin fabric.
  The breaking can be also created by another impurity; in this case the
local
  spirals of two impurities lock each other.

4)The bound state parity breaking in the t-J model is very similar
  to the parity breaking within the $2s_{1/2}-2p_{1/2}$ parity doublet of the 
  hydrogen atom.

\section{Acknowledgements}
Important discussions with  C. Batista and A. Sandvik
are acknowledged.


\begin{thebibliography}{99}
\bibitem{com1} Eq. (\ref{chiC}) is exact solution of the Coulomb problem
only in the isotropic case, $\beta_1=\beta_2$. In the general anisotropic 
case Eq. (\ref{chiC}) strictly speaking is valid only asymptotically.
\end{thebibliography}
\end{document}